\documentclass[conference]{IEEEtran}
\IEEEoverridecommandlockouts
\usepackage{cite}
\usepackage{amsmath,amssymb,amsfonts}
\usepackage{multirow}  % Align for multi rows of table
\usepackage{array}
\usepackage{booktabs} % For professional looking tables

\usepackage{fancyhdr}
\usepackage{graphicx}
\usepackage{textcomp}
\usepackage{xcolor}
\usepackage{algorithm} 
\usepackage{amssymb}
\usepackage{algpseudocode}
\usepackage{accents}
\def\BibTeX{{\rm B\kern-.05em{\sc i\kern-.025em b}\kern-.08em
    T\kern-.1667em\lower.7ex\hbox{E}\kern-.125emX}}
\begin{document}

\title{
%Influence of computation Granularity on Equitable Coordination in Multi-agent Power Systems
Equitable Coordination in Multi-agent Power Systems: Impacts of Computation Granularity
\thanks{This work is supported by the National Science Foundation (\# 2313768).}
}
% 1\textsuperscript{st}

% one block layout
\author{\IEEEauthorblockN{Yuhan Du, Javad Mohammadi}
\IEEEauthorblockA{
% \textit{Department of Civil, Architectural and Environmental Engineering} \\
\textit{University of Texas at Austin} \\
Austin, TX, USA \\
\{yuhandu, javadm\}@utexas.edu}
}

\maketitle
\begin{abstract} \label{s:0}
% Increasing penetration of distributed energy resources is fueling the evolution of our centralized electric grid to a multi-agent system. 
% System-level performance of multi-agent networks greatly depends on the communication and computation capabilities of nodes (customers). 
% The granularity of multi-agent networks indicates the size of agents.
% Its effect on equitable power coordination for customers affected by limited communication bandwidth (e.g., caused by sporadic internet access) is not studied extensively. 
% This paper investigates granularity in the context of agent-based power systems and studies its effect on equitable coordination. 
% The case studies leverage the \textit{Consensus + Innovations} approach to simulate the behavior of a multi-agent power system. 
The growing integration of distributed energy resources drives the centralized power system towards a decentralized multi-agent network. 
Operating multi-agent networks significantly relies on inter-agent communications. 
Computation granularity in this context refers to the number of nodes overseen by an agent. 
The impact of granularity on equitable power coordination, particularly among marginalized customers with limited communication bandwidth (e.g., intermittent internet connectivity) is not well studied. 
This work explores different levels of computation granularity for agent-based energy dispatch and studies their impact on equitable coordination. We will leverage and utilize the \textit{Consensus + Innovations} approach to model the equitable coordination of a multi-agent power system.

\end{abstract}


\section*{Nomenclature}

\begin{table}[htbp!]
\small
\begin{tabular}{lp{6.5cm}}

$i$& Agent index \\

$N_A$& Number of power system nodes (agents)\\

$A$& Set of agents in the power system\\

$c_{1, i}$, $c_{2, i}$& Generators cost function parameters\\

$P_{i}$& Electric output of agent $i$\\

$L_{i}$& Electric demand of agent $i$\\

$\underline{P}_{i}$, $\overline{P}_{i}$& Agent $i$'s minimum/maximum output limit\\

$\Omega_{\underline{B}}$& Set of generators reaching minimum generation limit in the energy aggregation\\

$\Omega_{\overline{B}}$& Set of generators reaching maximum generation limit in the energy aggregation\\

% $i\notin {\Omega_{\overline{B}}\cup\Omega_{\underline{B}}}$& denotes non-binding generator $i$\\
$\lambda_i^t$& Lagrangian multiplier of agent $i$ at iteration $t$\\

$P_{i}^t$& Electric output of agent $i$ at iteration $t$\\

$\lambda_i^*$& Lagrangian multiplier of agent $i$ at the optimal point\\

% $(P_{i}^{0})^{*}$& Optimal electric output of agent $i$ at previous iteration\\

% $L_{i}^0$& Electric demand of agent $i$ at previous iteration\\

$\beta_i^{t+1}$, $\alpha_i^{t+1}$& Tuning parameters of agent $i$ at iteration $t+1$\\

$\Omega_i$& Set of agent $i$'s neighbors \\

$N_m$& Maximum number of iterations for \textit{Consensus + Innovations} process \\

$f$& Objective function value \\
$f^*$& Optimal objective function value

\end{tabular}
\end{table}

\vspace{-0.5cm}

% *** I. Introduction ***
\section{Introduction} \label{s:1}
The complexity of electrical power systems is rising with the increased incorporation of distributed energy resources (DERs) \cite{li2023learning, mohammadi2023towards, mohammadi2023strategizing}. 
Growing penetration of smart devices with capabilities for communication and computation enables a transition towards decentralized information processing \cite{du2022learning}. 
This evolution highlights the importance of coordination among various entities of multi-agent systems \cite{mohammadi2016agent}.
%, effectively managing extensive distributed data and computation needs \cite{biagioni2020learning, mohammadi2016fully}. 
The scalability, privacy benefits, and robustness of these multi-agent information processing techniques have attracted much research interest \cite{li2023machine}. 
Within this framework, an agent is defined as a single node or a cluster of nodes that can perform local data processing and engage in inter-agent communication, thereby handling tasks collaboratively \cite{du2023need}. 
Multi-agent optimization methods establish a cooperative network among agents to solve complex problems by leveraging local computations and communications, reducing computation complexity \cite{du2022learning}.

The shift towards a more distributed energy system aggravates energy equity challenges \cite{van2021energy}. 
Energy equity (energy justice) encompasses a wide range of concerns including
%concerned with achieving equity in the social and economic participation of the energy system, 
%particularly crucial for marginalized communities to improve 
energy access \cite{bednar2020recognition}, affordability \cite{baker2021perspective, romero2021energy}, environmental sustainability \cite{shakya2023environmental}, threats of extreme weather\cite{castellanos2023synthesis}, availability of smart technologies \cite{tarekegne2021review} and equitable coordination \cite{tietjen2023women, ravikumar2023enabling}. 
% Recent research has redirected the focus of equity metrics from utility-scale to community-scale to focus more on marginalized agents \cite{parker2023observations}, which suffer from disconnection problems \cite{jenkins2021methodologies}.
These access limitations \cite{parker2023observations, jenkins2021methodologies} may directly impact the implementation of decision-making paradigms that are often modeled as multi-agent protocols.
In a multi-agent system, communication between agents is crucial to the performance of distributed optimization methods \cite{mohammadi2014distributed}. 
Sporadic and weak network connections can enlarge the energy equity gap, especially impacting marginalized agents of distributed multi-agent energy systems.
% \cite{baker2023metrics}. 

In our earlier work \cite{du2023need}, we discussed the need to study equity considerations in the context of multi-agent power systems. In the present work, we will explore the implications of computation granularity on energy equity of multi-agent systems. 
Computation granularity is generally defined as the level of detail in modeling time, space, and economic factors during decision-making processes \cite{sarfarazi2023improving, plazas2022national}. 
Cao et al. describe the granularity gap as the deficiencies that emerge from the trade-off between computational capabilities and the desired model granularity \cite{cao2021bridging}. 
Computation granularity is often regarded as the distributedness of the underlying multi-agent algorithm and the number of nodes/entities controlled/overseen by an agent. While the impact of computation granularity on multi-agent energy management is well-studied, equity implications are widely overlooked.
%In a multi-agent system, granularity denotes the size of individual agents during model simulations.

%Despite the extensive research covering various aspects of granularity, there appears to be a lack of depth in studies specifically related to the simulation of different computation granularities for achieving equitable coordination in energy aggregation problems, to the best of the author's knowledge.

To this end, this work seeks to model the trade-off associated with the computation granularity levels in the context of the multi-agent energy aggregation problem in the face of sporadic communications. 
%Our multi-agent decision is formulated as a consensus-based optimization problem. 
%In particular, 
We will use the \textit{Consensus + Innovations} approach \cite{kar2012distributed, du2023need} to develop our multi-agent framework where the \textit{consensus} term guides inter-agent agreements and local constraints are enforced through the \textit{Innovations} term.
This method has been broadly used to find distributed solutions for energy management problems. Examples include Economic Dispatch (ED) \cite{kar2014distributed} and DC Optimal Power Flow (DCOPF) \cite{mohammadi2014distributed}.
While the discussions in this paper are centered on resolving aggregation problems, the findings are also applicable to a broader understanding of energy equity across a spectrum of multi-agent energy optimization problems. 

% The remainder of the paper is organized in the following manner: 
% Section II presents the mathematical formulations for the aggregation problem and the \textit{Consensus + Innovations} approach, as well as the models for disruption and granularity. 
% Section III illustrated the simulation parameters.
% The IEEE 123 test feeder is utilized to compare the results of five levels of granularities under sporadic communication lines during both normal operations and disrupted scenarios. 
% Section IV displayed the simulation results.
% The relationship between the granularity and the converging rate is discussed.
% The paper is concluded in Section V.

% *** II. Mathematical Models ***
% \vspace{0.1in}
\section{Mathematical Models}  \label{s:2}
This section presents the mathematical foundation of the energy aggregation problem and the \textit{Consensus + Innovations} multi-agent solution approach. We will also discuss the granularity and disruption models.
%Modeling of disruption and granularity are included to demonstrate the impact of granularity on equitable coordination.

\subsection{Energy Aggregation Problem} \label{ss:Model_ea}

This work employs a multi-agent perspective of the electrical grid, wherein each agent represents a power system node (or a cluster of nodes) that may be capable of power consumption, production, or both. 
%The electric connections among these agents facilitate the delivery of power and form the foundational power distribution network. 
%
The fundamental function of every electric network is to distribute energy in the most cost-effective manner. 
%Building upon these initial concepts, the aggregation problem targets to reduce the aggregate systematic energy cost while maintaining the balance between supply and demand and adhering to the physical constraints of agents' resources.
The dispatch cost is modeled as the aggregate energy production cost. The centralized representation of this problem can be presented as \eqref{eq:ea1}-\eqref{eq:ea3}.

\vspace{-0.5cm}
\small
\begin{align}
    \min_{P_{i}} \quad & \sum_{i=1}^{N_A}C_i(P_{i})=\sum_{i=1}^{N_A}(c_{1, i} P_{i}^2+c_{2, i} P_{i}) \label{eq:ea1} \\
     \textrm{s.t.} \quad & \sum_{i=1}^{N_A}P_{i}=\sum_{i=1}^{N_A}L_{i} \label{eq:ea2} \\
                         & \underline{P}_{i} \leq P_{i}\leq \overline{P}_{i} \label{eq:ea3}
\end{align}
\normalsize

The Lagrange multiplier associated with \eqref{eq:ea2} is interpreted as the price signal ($\lambda$), uniform among all agents that have not reached the limits of \eqref{eq:ea3}. 
To find a solution for the first-order optimality condition of \eqref{eq:ea1}-\eqref{eq:ea3}, the inequality constraints on generation are initially omitted. 
Consequently, the Lagrangian of the resulting optimization problem simplifies to \eqref{eq:lag_multi}. 
The first-order optimality conditions of \eqref{eq:lag_multi} are presented by \eqref{eq:fooc1} and \eqref{eq:fooc2}.

\vspace{-0.5cm}
\small
\begin{align} \label{eq:lag_multi}
    \mathcal{L} = \sum_{i=1}^{N_A}C_i(P_{i}) + \lambda \cdot \left(\sum_{i=1}^{N_A}L_{i} - \sum_{i=1}^{N_A}P_{i}\right)
\end{align}
\normalsize

\vspace{-0.5cm}
\small
\begin{align}
    \frac{\partial\mathcal{L}}{\partial P_{i}} = 2c_{1, i} P_{i} + c_{2, i} - \lambda = 0 \label{eq:fooc1} \\
    \frac{\partial\mathcal{L}}{\partial \lambda}=\sum_{i=1}^{N_A}L_{i} - \sum_{i=1}^{N_A}P_{i} = 0 \label{eq:fooc2}
\end{align}
\normalsize

Let us assume that $\lambda^*$ and $P_{i}^*$ denote the solutions to \eqref{eq:fooc1} and \eqref{eq:fooc2}. Then, these solutions can be derived through \eqref{eq:lamStar} and \eqref{eq:fooc5}. 
In this context, $\lambda^*$ as outlined in \eqref{eq:lamStar} denotes the Lagrange multiplier relevant to agents whose inequality constraints are not binding. 
The term ${\Omega_{\overline{B}}\cup\Omega_{\underline{B}}}$ indicates the set comprising non-binding variables. 
Furthermore, equations \eqref{eq:fooc3}-\eqref{eq:fooc5} specifies the optimal power generation $P_{i}^*$ for agents operating under both non-binding and binding scenarios. 
Consequently, equation \eqref{eq:lamStar} can be revised and presented as \eqref{eq:lamStar2}.

\vspace{-0.5cm}
\small
\begin{align}
    \lambda^* = \left(\sum_{i=1}^{N_A}\frac{1}{2c_{1, i}}\right)^{-1} \left(\sum_{i=1}^{N_A}L_{i} + \sum_{i=1}^{N_A}\frac{c_{2, i}}{2c_{1, i}}\right) \label{eq:lamStar}
\end{align}
\normalsize

\vspace{-0.5cm}
\small
\begin{align}
    2c_{1, i} P_{i}^* + c_{2, i} - \lambda^* = 0, i\notin {\Omega_{\overline{B}}\cup\Omega_{\underline{B}}} \label{eq:fooc3} \\
    P_{i}^* = \overline P_{i}, i\in {\Omega_{\overline{B}}} \label{eq:fooc4} \\
    P_{i}^* = \underline P_{i}, i\in {\Omega_{\underline{B}}} \label{eq:fooc5}
\end{align}
\normalsize

\vspace{-0.5cm}
\small
\begin{multline} \label{eq:lamStar2}
    \lambda^* = \left(\sum_{i\notin {\Omega_{\overline{B}}\cup\Omega_{\underline{B}}}}\frac{1}{2c_{1, i}}\right)^{-1} \left[\sum_{i=1}^{N_A}L_{i} - \sum_{i\in {\Omega_{\overline{B}}}} \overline P_{i} \right. \\
    \left. \quad - \sum_{i\in {\Omega_{\underline{B}}}} \underline P_{i} + \sum_{i\notin{\Omega_{\overline{B}}\cup\Omega_{\underline{B}}}}\frac{c_{2, i}}{2c_{1, i}} \right]
\end{multline}
\normalsize

Thus, $\lambda^*$ and $P_{i}^*$ can be analytically solved for a centralized energy aggregation problem through \eqref{eq:fooc3}-\eqref{eq:lamStar2}.

\subsection{Consensus + Innovations Approach} \label{ss:Model_cpi}

The \textit{Consensus + Innovations} approach aims to solve the energy aggregation problem through a distributed collaboration \cite{kar2014distributed}. 
In this iterative approach, agents perform local computations and exchange information with neighboring agents. 
In order to find a distributed solution for  \eqref{eq:ea1}-\eqref{eq:ea3}, agents make individual copies of the ``consensus variable" (denoted as $\lambda$) and assign them to neighboring agents. 
Through cooperation, agents strive to reach an agreement on the values of $\lambda$ and determine the optimal generation level of each agent, thus moving towards fulfilling the supply-demand equilibrium locally.
Differently put, this strategy seeks to fulfill the optimality conditions of the energy aggregation optimization problem in a fully distributed manner.

The optimization variables are updated iteratively by using \eqref{eq:lam} and \eqref{eq:PG} update rules.
During each iteration, agent $i$ executes the updates independently and communicates the newly calculated values of $\lambda_i$ to neighboring agents. 
% The update of $\lambda$ in \eqref{eq:lam} comprises two components: the \textit{Consensus} component which seeks to achieve uniformity in the values of $\lambda$ among all agents, and the \textit{Innovations} component which is responsible for sustaining local constraints. 
 This update protocol converges to the optimal solution provided that the communication graph is connected and the hyperparameters $\alpha$ and $\beta$ are appropriately tuned \cite{kar2012distributed, mohammadi2016distributed}.
The updates resume until a mutual consensus on $\lambda$ and the optimality requirements of \eqref{eq:ea1}-\eqref{eq:ea3} are met. 

\vspace{-0.0cm}
\small
\begin{align} \label{eq:lam}
    \lambda_i^{t+1} = \lambda_i^t - \beta_i^{t+1} \sum_{j\in\Omega_i}(\lambda_i^t - \lambda_j^t) - \alpha_i^{t+1}(P_{i}^t-L_{i})
\end{align}
\vspace{-0.2cm}
\begin{align} \label{eq:PG}
P_{i}^{t+1}=
\begin{cases}
    \frac{\lambda_i^{t+1}-c_{2, i}}{2c_{1, i}},& 0\leq \frac{\lambda_i^{t+1}-c_{2, i}}{2c_{1, i}}\leq \overline{P}_{i}\\
    \overline{P}_{i},                   & \frac{\lambda_i^{t+1}-c_{2, i}}{2c_{1, i}}>\overline{P}_{i}\\
    0,                               &\frac{\lambda_i^{t+1}-c_{2, i}}{2c_{1, i}}<0
\end{cases}
\end{align} 
\normalsize

The convergence is quantified by measuring the distance between agents' local copies ($\lambda_i$) and the optimal $\lambda^*$, where $\lambda^*$ is derived by solving the centralized problem \eqref{eq:lamStar2}. 
This discrepancy is compared against a pre-determined threshold ($\varepsilon$) as in \eqref{eq:conv}.

\vspace{-0.5cm}
\small
\begin{align}
|\lambda_i-\lambda^{*}| \leq \varepsilon, \forall {i\in A} \label{eq:conv}
\end{align}
\normalsize

\subsection{Disruption \& Granularity Modeling} \label{ss:Model_dg}

In multi-agent cyber-physical systems, agents are interconnected by physical infrastructure (e.g. power lines) and communication means. 
In this paper, we assume that the communication network mirrors the physical connections. Put differently, agents communicate only with their physically connected neighbors.
%when they are physically connected, which is referred to as physics-based communication.
This paper investigates how different computation granularity levels affect equitable coordination across the electric network. 
In our recent work \cite{du2023need}, we studied the lack of access to reliable communication as a barrier for marginalized consumers (agents) to participate in energy aggregation and coordination. 
As outlined in Section \ref{ss:Model_cpi}, agents must exchange local $\lambda$ values with neighboring agents in each iteration to achieve convergence. 
However, if communications are unreliable, a struggling agent $i$ may not be able to send or receive the latest $\lambda_i$ to/from its neighboring agents in $\Omega_i$. 
This paper assumes that the agent continues to use the last communicated $\lambda$ value (received before the disruption) in subsequent iterations.

Computation granularity in the context of the multi-agent framework, refers to the number of nodes that are overseen by an agent in a $N_A$-node system.
% At the distribution level, agents operate in a fully distributed manner. 
In a fully distributed setup, each node is regarded as an individual agent, and the number of agents equals $N_A$.
However, nodes can be clustered under the supervision of a super-agent. These super-agents have access to the information of their nodes.
%These managers have a comprehensive awareness of the resources within their cluster of nodes. 
Superagents (clusters of nodes) can be partitioned according to geographical areas or the physical topology of the infrastructure
%, creating a cellular framework 
% \cite{eder2016resilience}. 
\cite{disfani2015sdp}.
In the case where a superagent's area contains all $N_A$ nodes, the setup transforms into a centralized architecture.

% *** III. Simulation Setup ***
% \vspace{0.1in}
\section{Simulation Setup}  \label{s:simulation}
The algorithm is evaluated on a modified version of the IEEE 123-node test system \cite{schneider2017analytic}.
%, which typically indicates a distribution-level, root-like network configuration.
The modifications include:
(i) exclusion of transformers, 
(ii) replacing three-phase switches with physical and communication connections between nodes 13-152, 18-135, 54-94, 60-160, 97-197, and 151-300,
%, also operated as communication lines in a physics-based system,
% (iii) conversion to a single-phase framework with integration of three-phase loads, 
% (iv) application of a DC approximation assumption, 
(iii) incorporation of five additional power sources located at nodes 1, 35, 60, 76, and 144. 
% The IEEE 123-node test feeder typically indicates a distribution-level, root-like network configuration.

To model communication disruptions without causing any islanding, two communication lines (between nodes 54-94 and 151-300) are selectively interrupted. % without losing generality. 
These disruptions are implemented by disrupting the connections at a designated iteration during the \textit{Consensus + Innovations} update cycle and reconnecting them at a subsequent iteration. 
Following our previous work \cite{du2023need}, one particular scenario of communication disruption is examined, wherein lines 54-94 and 151-300 experience an interruption at the 20th iteration and are reconnected at the 400th iteration. 
During this interruption, agents continue using the last communicated value of $\lambda$ for the succeeding iterations. 
A scenario without disruptions serves as a baseline for comparison and verifying the convergence behavior.

\begin{figure}[htbp]
\centering
\setlength{\abovecaptionskip}{0.cm}
\vspace{-.3cm}
\includegraphics[width=0.7\columnwidth]{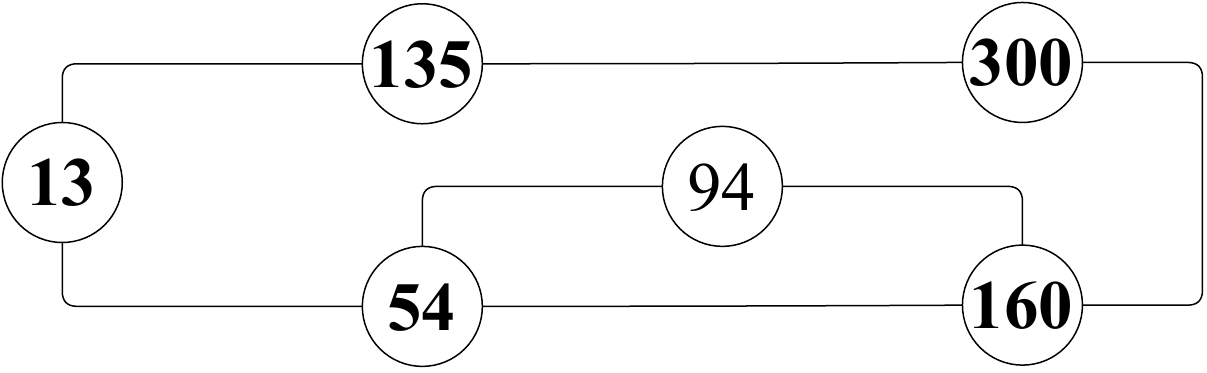} % 0.47
\caption{Network topology of the 6 superagents that represent the 128-node test system.
Circles denote superagents and the number of the node that is hosting the superagent.
%with the internal number representing the corresponding node of the superagent manager. 
Bold numbers represent superagents that own generations.}
\label{f:6agent}
\vspace{-.3cm}
\end{figure}

\begin{figure}[htbp]
\centering
\setlength{\abovecaptionskip}{0.cm}
\vspace{-.3cm}
\includegraphics[width=0.7\columnwidth]{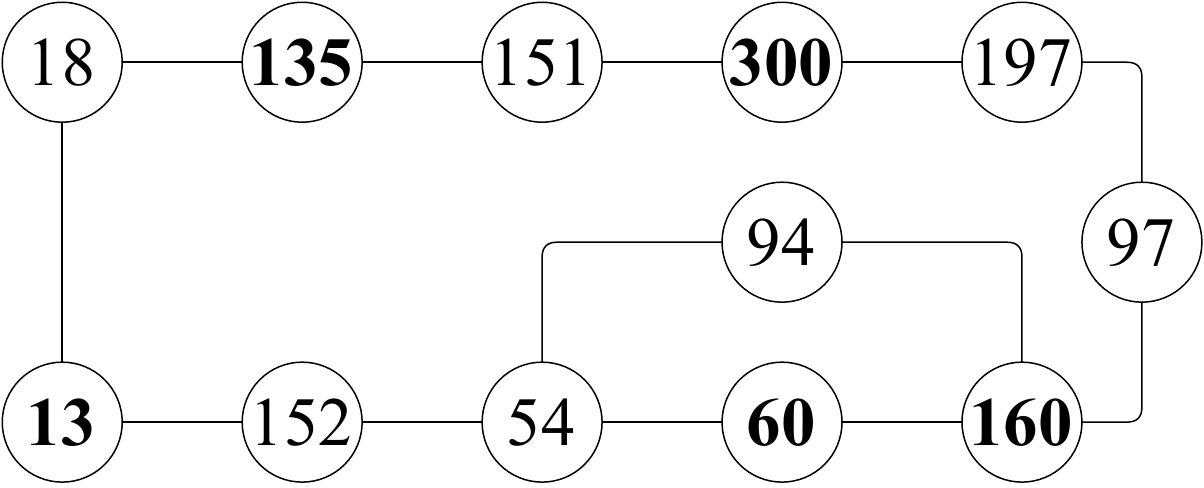} % 0.47
\caption{Network topology of the 12 superagents that represent the 128-node test system.
Circles denote superagents and the number of the node that is hosting the superagent.
%with the internal number representing the corresponding node of the superagent manager. 
Bold numbers represent superagents with generators.}
\label{f:12agent}
\vspace{-.3cm}
\end{figure}

\begin{figure}[htbp]
\centering
\setlength{\abovecaptionskip}{0.cm}
\vspace{-.3cm}
\includegraphics[width=0.7\columnwidth]{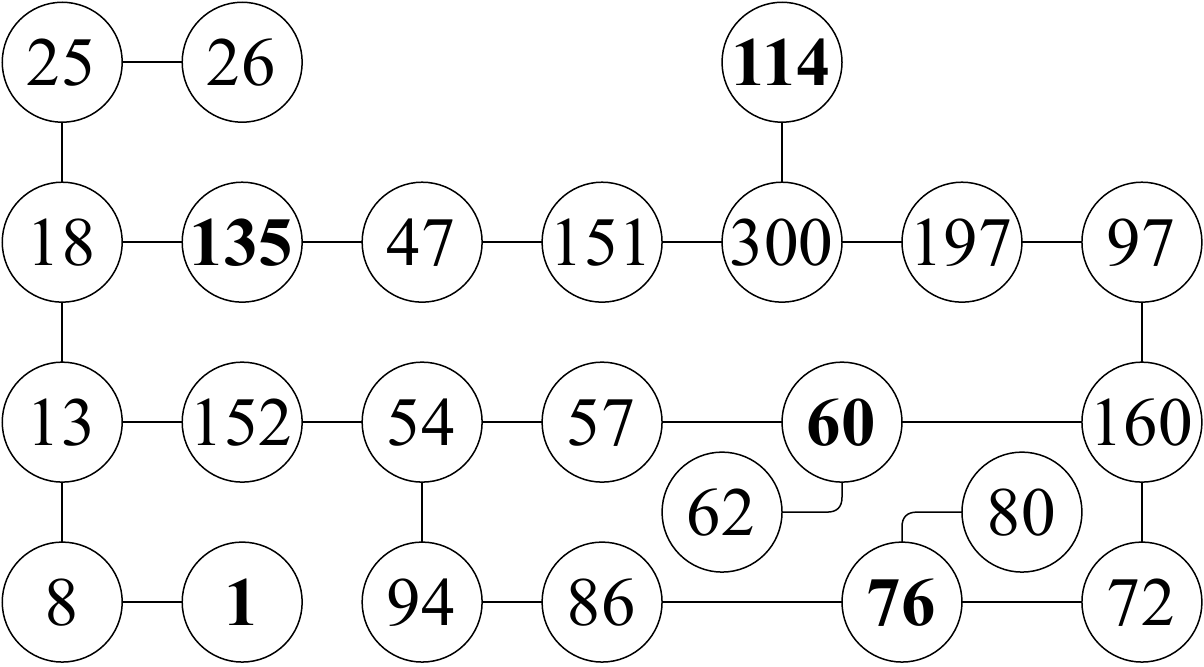} % 0.47
\caption{Network topology of the 24 superagents that represent the 128-node test system.
Circles denote superagents and the number of the node that is hosting the superagent.
%with the internal number representing the corresponding node of the superagent manager. 
Bold numbers represent superagents with generators.
}
\label{f:24agent}
\vspace{-.3cm}
\end{figure}

\begin{figure}[htbp]
\centering
\setlength{\abovecaptionskip}{0.cm}
\vspace{-.3cm}
\includegraphics[width=0.7\columnwidth]{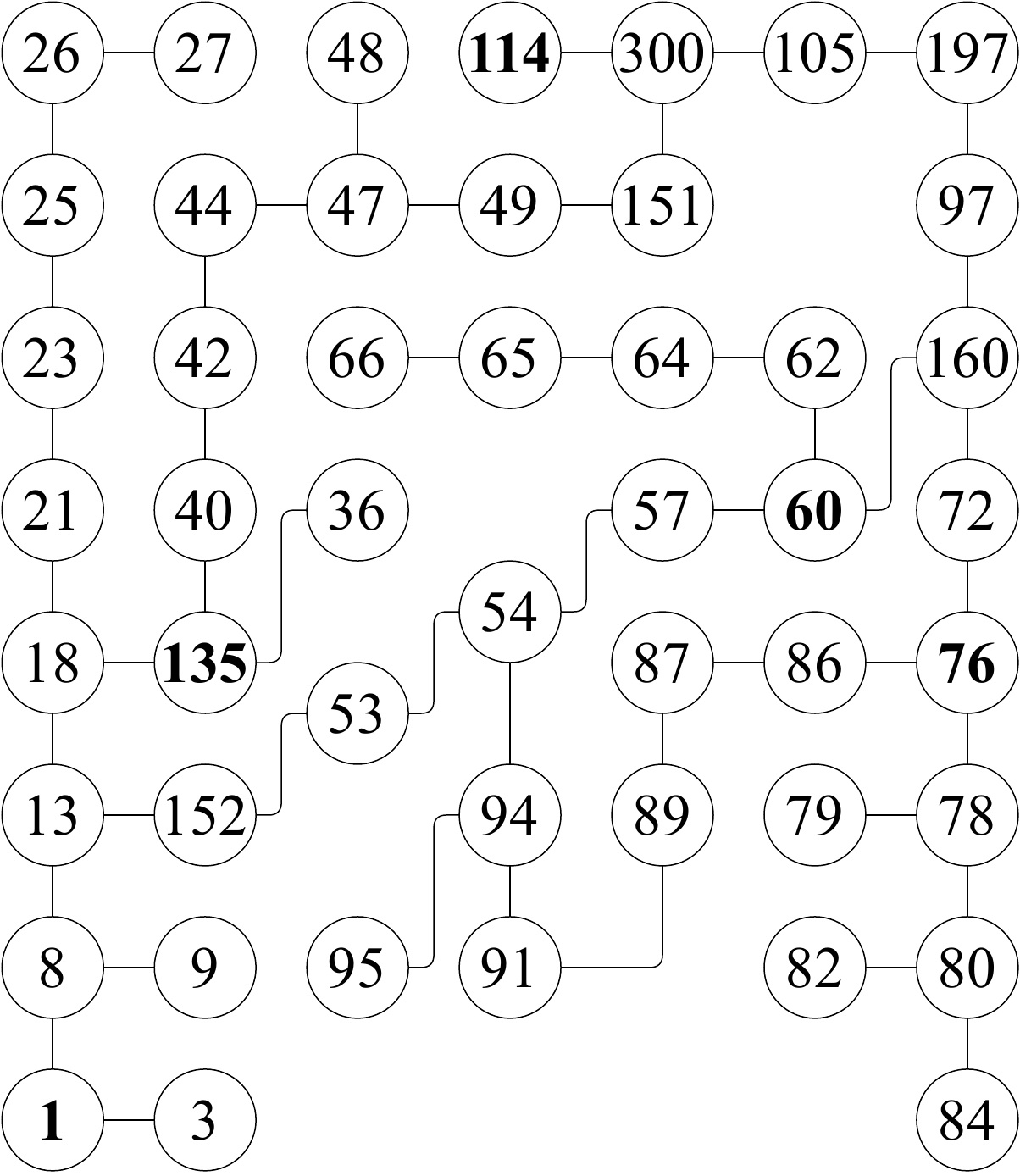} % 0.47
\caption{Network topology of the 48 superagents that represent the 128-node test system.
Circles denote superagents and the number of the node that is hosting the superagent.
%with the internal number representing the corresponding node of the superagent manager. 
Bold numbers represent superagents that own generations.
}
\label{f:48agent}
\vspace{-.0cm}
\end{figure}

To study the effect of computation granularity, we consider five different levels of granularity: clustering nodes into (i) a 6-agent, (ii) a 12-agent, (iii) a 24-agent, (iv) a 48-agent, and (v) a 123-agent network. Figures \ref{f:6agent}-\ref{f:48agent} illustrate the inter-super-agent connections for different numbers of super-agents. Each node is considered an independent agent in the most granular case, that is the 123-agent network.

The convergence criterion for the distributed update process is established with a threshold of $\varepsilon = 0.005$ (see \eqref{eq:conv}). 

The iterative process is capped at different maximum iterations for different granularities: $N_m = 1000$ for the 6-agent network, $N_m = 2000$ for both the 12-agent and 24-agent networks, and $N_m = 3000$ for the 48-agent and 123-agent networks. 
These simulations are conducted using the PyCharm IDE (Version 2023.2.5) within a Python 3.6 environment, operated on a MacBook Pro (Apple M3 Max, 2023).

% *** IV. Results ***
% \vspace{0.1in}
\section{Results}  \label{s:results}
\begin{figure}[htbp]
\centering
\setlength{\abovecaptionskip}{0.cm}
\vspace{-.3cm}
\includegraphics[width=0.85\columnwidth]{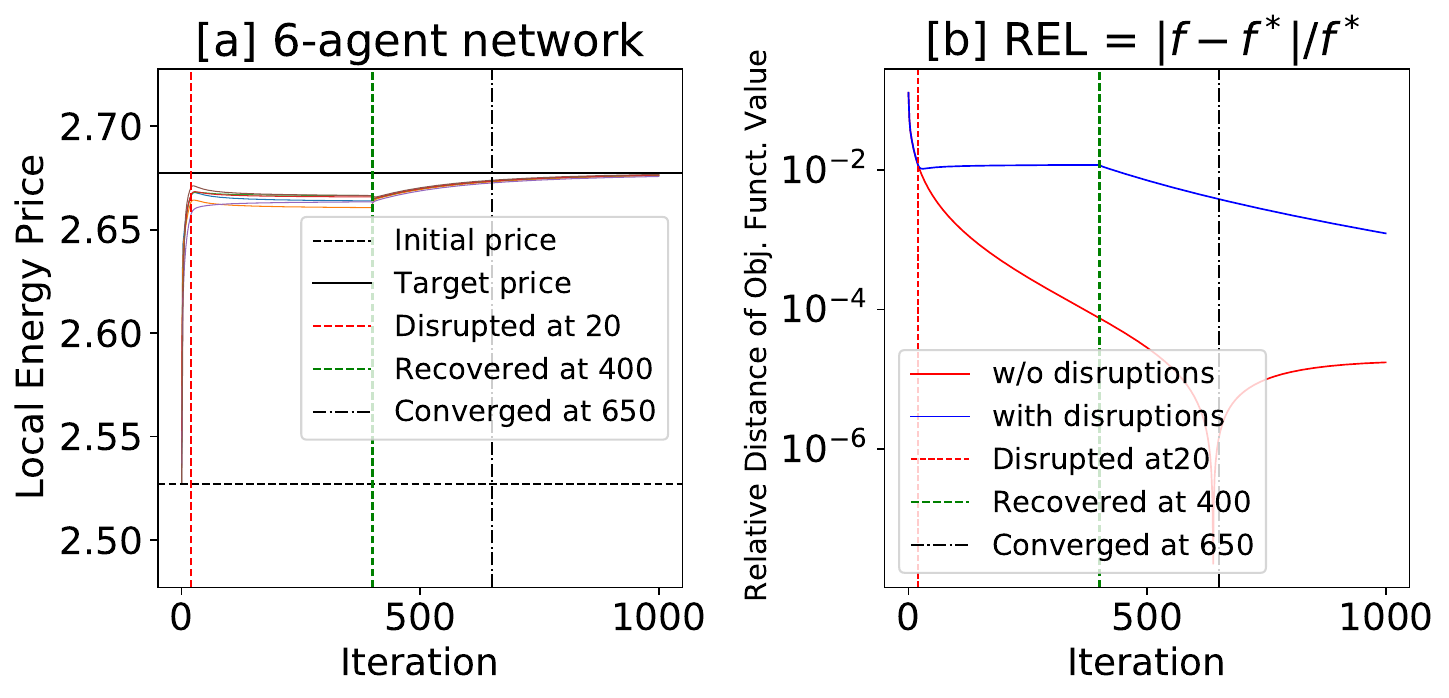}
\caption{Convergence performance for the scenario with disruptions under the 6-agent network. 
[a] shows $\lambda$ of all agents throughout iterations. 
[b] presents the relative distance of the objective function value $REL = {|f - f^*|}/{f^*}$.
}
\label{f:6_s1}
\vspace{-.3cm}
\end{figure}

\begin{figure}[htbp]
\centering
\setlength{\abovecaptionskip}{0.cm}
\vspace{-.3cm}
\includegraphics[width=0.85\columnwidth]{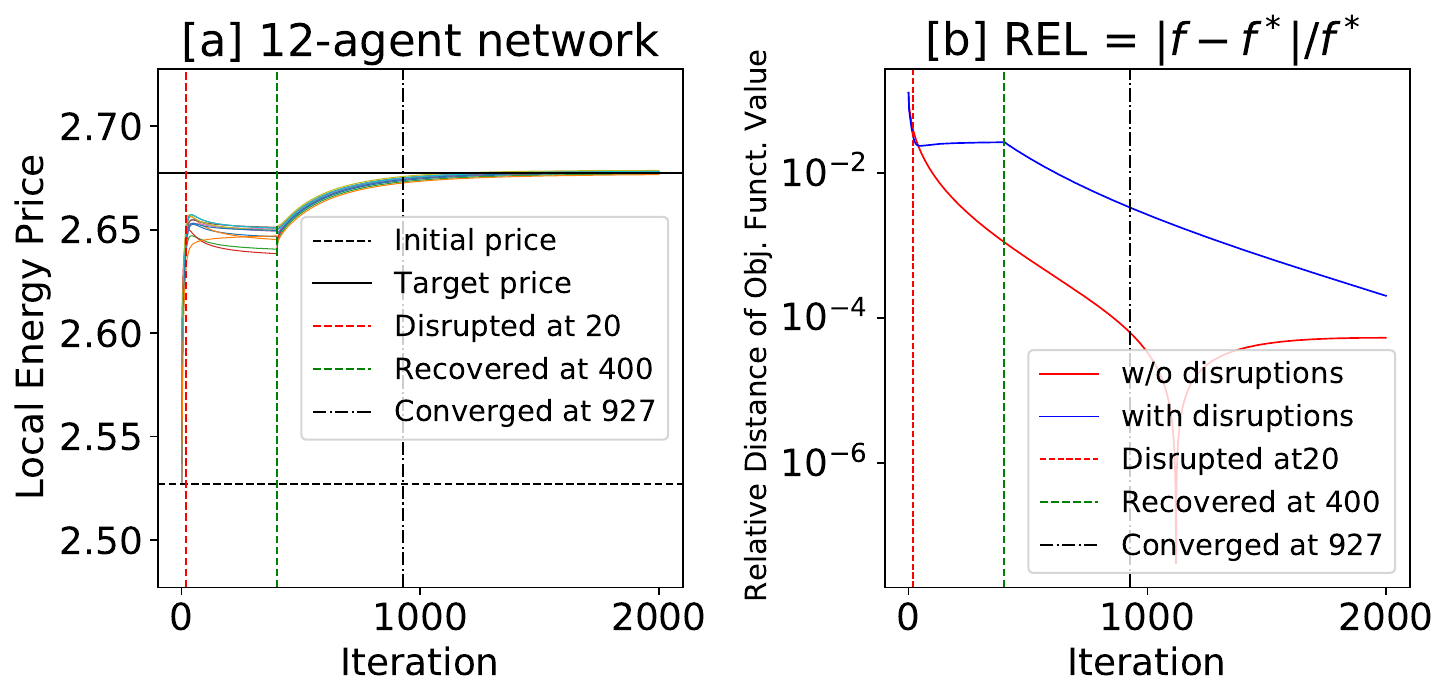}
\caption{Convergence performance for the scenario with disruptions under the 12-agent network. 
[a] shows $\lambda$ of all agents throughout iterations. 
[b] presents the relative distance of the objective function value $REL = {|f - f^*|}/{f^*}$.
}
\label{f:12_s1}
\vspace{-.3cm}
\end{figure}

\begin{figure}[htbp]
\centering
\setlength{\abovecaptionskip}{0.cm}
\vspace{-.3cm}
\includegraphics[width=0.85\columnwidth]{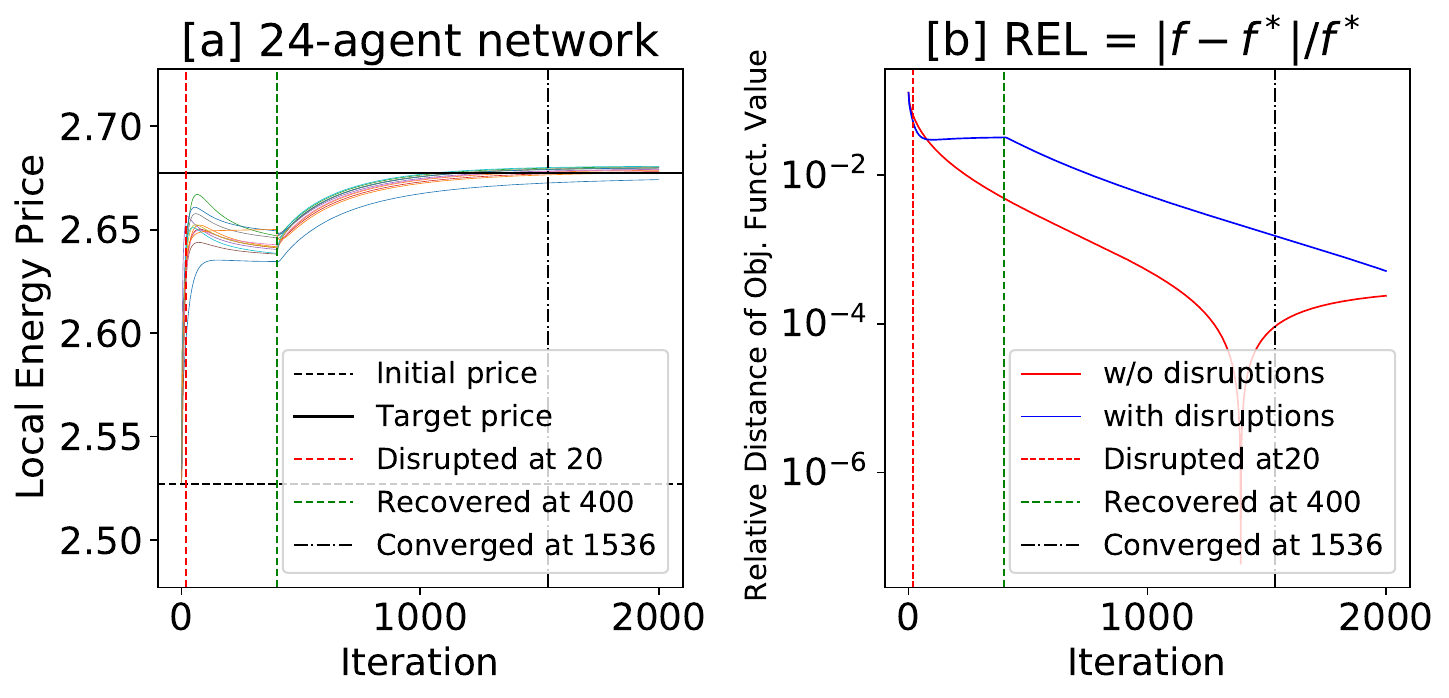}
\caption{Convergence performance for the scenario with disruptions under the 24-agent network. 
[a] shows $\lambda$ of all agents throughout iterations. 
[b] presents the relative distance of the objective function value $REL = {|f - f^*|}/{f^*}$.
}
\label{f:24_s1}
\vspace{-.3cm}
\end{figure}

\begin{figure}[htbp]
\centering
\setlength{\abovecaptionskip}{0.cm}
\vspace{-.3cm}
\includegraphics[width=0.85\columnwidth]{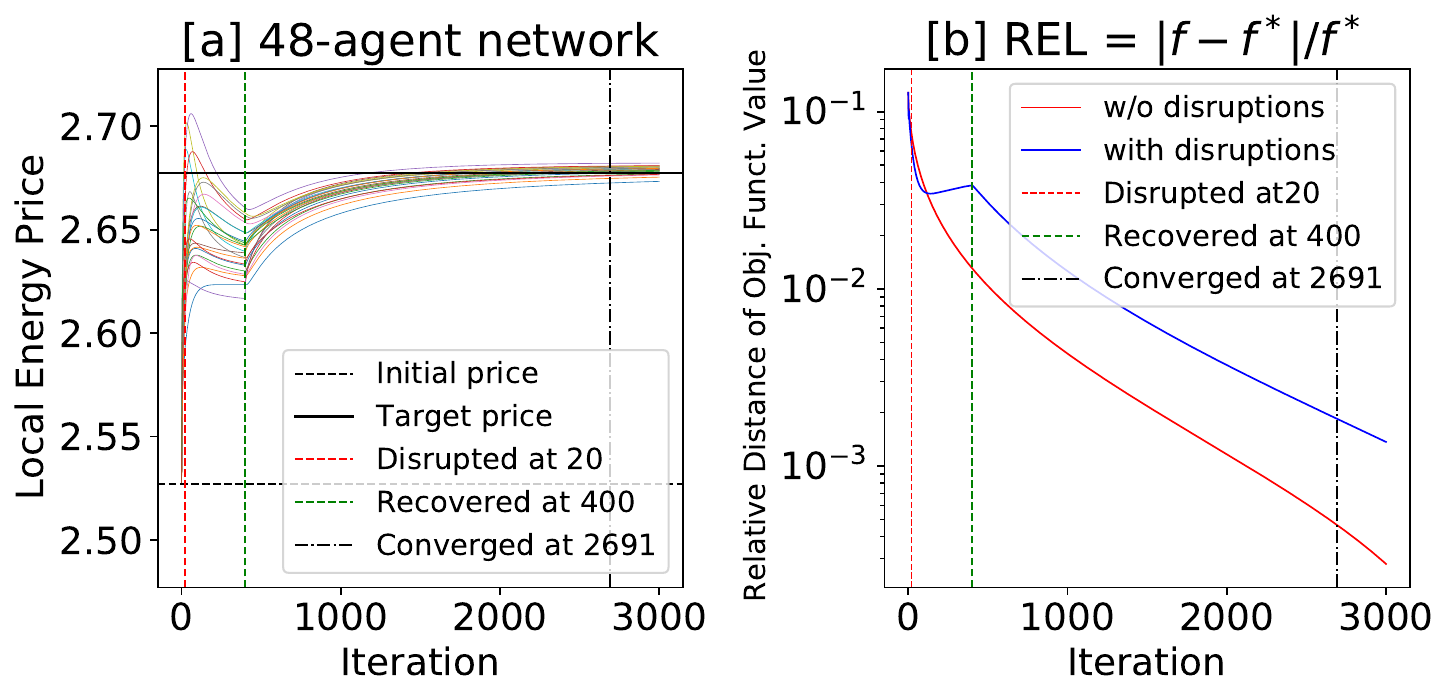}
\caption{Convergence performance for the scenario with disruptions under the 48-agent network. 
[a] shows $\lambda$ of all agents throughout iterations. 
[b] presents the relative distance of the objective function value $REL = {|f - f^*|}/{f^*}$.
}
\label{f:48_s1}
\vspace{-.15cm}
\end{figure}

\begin{figure}[htbp]
\centering
\setlength{\abovecaptionskip}{0.cm}
\vspace{-.2cm}
\includegraphics[width=0.85\columnwidth]{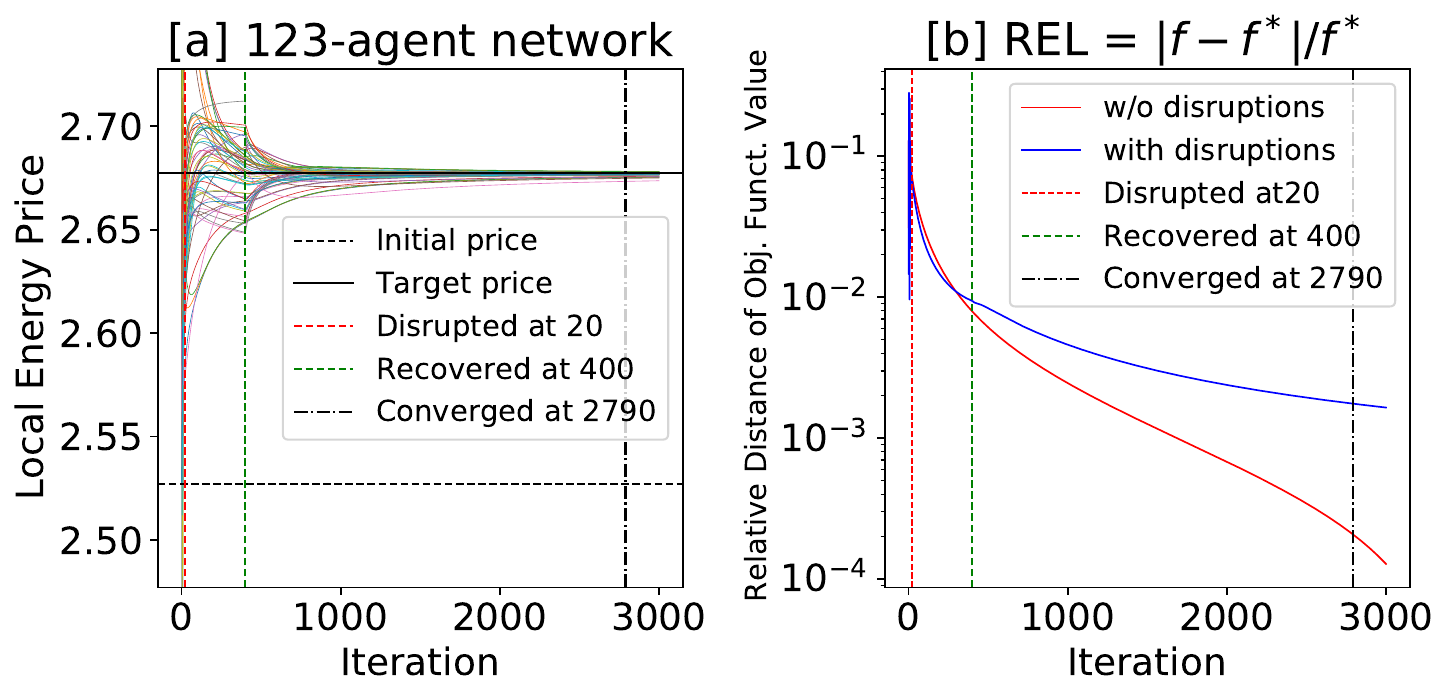}
\caption{Convergence performance for the scenario with disruptions under the 123-agent network. 
[a] shows $\lambda$ of all agents throughout iterations. 
[b] presents the relative distance of the objective function value $REL = {|f - f^*|}/{f^*}$.
}
\label{f:123_s1}
\vspace{-.3cm}
\end{figure}

The convergence performances for all five levels of granularities are presented in Fig. \ref{f:6_s1} through \ref{f:123_s1}. 
During disruptions in communication, which are initiated at the 20th iteration and denoted by a red dotted line in [a], agents continue to use the last communicated $\lambda$ value in subsequent iterations. 
This leads to a divergence in consensus from the accurate value in [a], where $\lambda$ stops to approach the target value, and in [b], where the relative distance of the objective function (REL) suspends its decline. 
It is only upon the recovery of communication at the 400th iteration that the error resumes declining at a rate correlated with the network's granularity.

\begin{figure}[htbp]
\centering
\setlength{\abovecaptionskip}{0.cm}
\vspace{-.3cm}
\includegraphics[width=0.7\columnwidth]{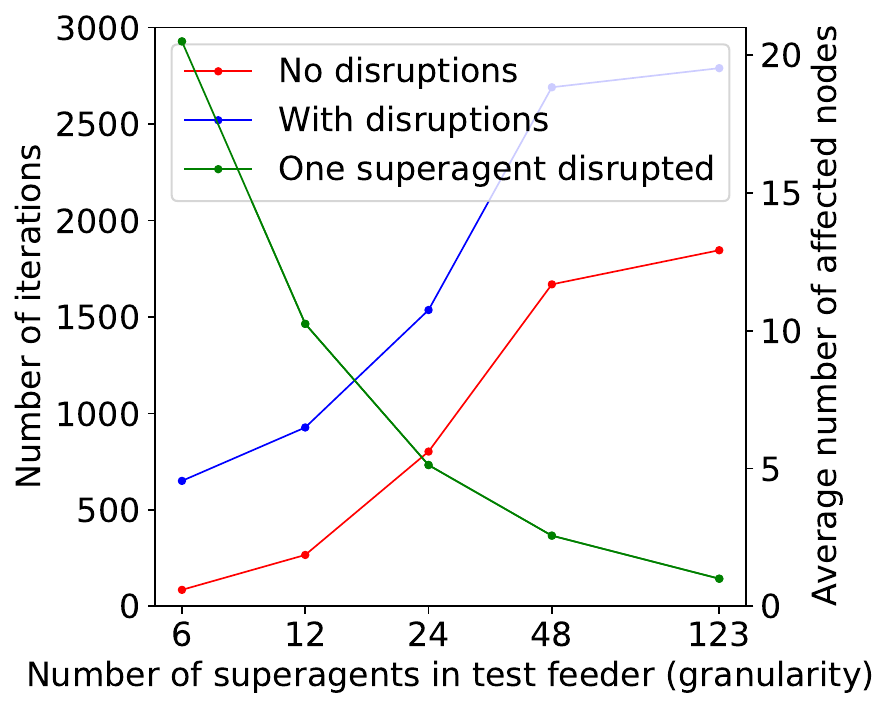}
\caption{Relationship between the number of superagents in the IEEE 123-node test feeder (granularity level) and the number of iterations required for convergence under two scenarios (blue and red curve, left y-axis). The green curve presents the average number of affected nodes under each granularity level (right y-axis).
}
\label{f:gvi}
\vspace{-.3cm}
\end{figure}

% The number of iterations required for convergence for the undisrupted and disrupted scenario%, which provides a baseline comparison, 
% is summarized in Table \ref{t:results}. 
%Each network configuration achieves successful convergence to the stipulated target energy price $\lambda^*$.

% \begin{figure}[htbp]
% \centering
% \setlength{\abovecaptionskip}{0.cm}
% \vspace{-.3cm}
% \includegraphics[width=0.8\columnwidth]{Figures/123_s0.pdf}
% \caption{Convergence performance for the scenario with no disruptions under the 123-agent network. 
% % The vertical dashed line indicates the iteration that the convergence criterion is met. 
% % The horizontal dashed line shows the initial value (initial price). 
% % The horizontal solid line shows the optimal value. 
% [a] shows $\lambda$ of all agents throughout iterations. 
% [b] presents the relative distance of the objective function value $REL = {|f - f^*|}/{f^*}$.
% }
% \label{f:123_s0}
% \vspace{-.3cm}
% \end{figure}

The number of iterations required to satisfy the convergence criterion (see \eqref{eq:conv}) with and without disruptions are listed in Table \ref{t:results} and visually illustrated in Figure \ref{f:gvi}.

\begin{table}[htbp]
\centering
\vspace{-.3cm}
\caption{Number of Iterations Until Convergence for Five Levels of Computation Granularities.}
\label{t:results}
\begin{tabular*}{0.4\textwidth}{@{\extracolsep\fill}llll@{\extracolsep\fill}}
\toprule
\textbf{Granularity} & \textbf{No Disruptions}      & \textbf{With Disruption}\\
\midrule
6-agent & 85 & 650 \\ 
% \midrule
12-agent & 266 & 927 \\ 
% \midrule
24-agent & 802 & 1536 \\ 
% \midrule
48-agent & 1669 & 2691 \\ 
% \midrule
123-agent & 1846 & 2790 \\ 
\bottomrule
\end{tabular*}
\vspace{-.2cm}
\end{table}

Based on Fig. \ref{f:gvi}, systems with more agents require more computational resources (higher iteration counts) to achieve convergence, as the red curve indicates.

For instance, a system comprising 6 agents may converge in fewer than 100 iterations, whereas systems with 48 or 123 agents typically require over 1500 iterations.
Communication disruptions, represented by the blue curve, extend the number of iterations needed for all system granularities to reach convergence. 
The disruption impact is reduced on smaller networks; the 6-agent and 12-agent systems experience an increase of under 700 iterations, whereas the 48-agent and 123-agent systems exceed 900 additional iterations.
Fewer superagents in a system benefit from a fast converging rate and more resilience to communication disruptions.
However, a system with fewer superagents (more nodes under one superagent) is more vulnerable to cyber-attacks and superagent-level failure.
The green curve displays the average number of affected nodes when one superagent is disrupted.
%More nodes are affected by a less distributed system during superagent disruption. 
The trade-off of computation granularity calls for an optimal choice of the number of superagents.
%Employing a rule of thumb, a network size approximating 24 agents (positioned at the inflection point of the blue and green curve) emerges as an optimal choice for granularity when encountering inequitable coordination scenarios.

% *** V. Conclusion ***
% \vspace{0.1in}
\section{Conclusion}  \label{s:4}

This study explores the concept of computation granularity within multi-agent power systems and studies its influences on equity of energy management.
Leveraging the \textit{Consensus + Innovations} framework, the research simulates scenarios where a multi-agent power system faces unreliable communication links. 
Findings from these simulations showcase the trade-off in computation granularity. 
A more distributed network (smaller superagents) is more sensitive to communication disruptions, whereas a less distributed network (larger superagents) is more vulnerable to superagent-level failures.
%The study further favors a rule of thumb to choose an optimal granularity that compromises equitable coordination and cyber security.

Recent studies on multi-agent systems often ignore the influence of computation granularity on equitable coordination in power systems. 
Addressing this gap, simulations of five levels of granularity are conducted using the IEEE 123-node test feeder. 
Future work will involve extended simulations on larger systems with randomly affected communication lines, to verify the most suitable level of granularity.

% *** VI. References ***
% \vspace{0.1in}
% \section{Acknowledgment}
% A part of this research was supported by ...

\bibliographystyle{IEEEtran}
\bibliography{conference.bib}

\end{document}